\documentclass[prl,aps,twocolumn,10pt]{revtex4-1}
\usepackage[utf8x]{inputenc}

\usepackage{amsmath,amssymb,graphicx,xcolor,esint}

\newcommand{\Caf}[4]{{\bf C}_{#1 #2 #3 #4}^{\left( \alpha \right)}}
\newcommand{\Cdf}[4]{{\bf C}_{#1 #2 #3 #4}^{\left( \delta \right)}}
\newcommand{\Cdpf}[4]{{\bf C}_{#1 #2 #3 #4}^{\left( \delta' \right)}}
\newcommand{\Cbf}[4]{{\bf C}_{#1 #2 #3 #4}^{\left( \beta \right)}}

\newcommand{\Ei}{{\bf E}_{i}}

\newcommand{\reshata}{\hat{r}_{nl}^{\left( \alpha \right)}}
\newcommand{\reshatb}{\hat{r}_{nl}^{\left( \beta \right)}}
\newcommand{\resresa}{r_{nl}^{\left( \alpha \right)}\left(x_{nl}^{\left(\alpha\right)} \right)}
\newcommand{\resresb}{r_{nl}^{\left( \beta \right)}\left(x_{nl}^{\left(\beta\right)} \right)}
\newcommand{\alphares}{\alpha_{nl} \left(x_{nl}^{\left(\alpha\right)} \right)}
\newcommand{\betares}
{\beta_{nl} \left(x_{nl}^{\left(\beta\right)} \right)}
\newcommand{\alphahat}{\hat{\alpha}_{nl}}
\newcommand{\betahat}{\hat{\beta}_{nl}}

\newcommand{\re}[1]{\mbox{Re}\left\{#1\right\}}

\newcommand{\im}[1]{\mbox{Im}\left\{#1\right\}}

\newif\ifpaper
\newif\ifpaperm

\papermfalse
\papertrue

\begin{document}

\title{On the electromagnetic resonances of nanoparticles}

\author{Carlo Forestiere}
\affiliation{ Department of Electrical Engineering and Information Technology, Universit\`{a} degli Studi di Napoli Federico II, via Claudio 21,
 Napoli, 80125, Italy}
\author{Giovanni Miano}
\affiliation{ Department of Electrical Engineering and Information Technology, Universit\`{a} degli Studi di Napoli Federico II, via Claudio 21,
 Napoli, 80125, Italy}

\begin{abstract}
A spectral technique is applied to evaluate the resonance frequencies of the full retarded scattering from spherical nanoparticles. This approach allows one to unambiguously identify the modes that are responsible of both the peaks and the asymmetric lineshapes of the scattered power spectra. The fundamental differences between the electromagnetic scattering properties of dielectric and metal spheres are discussed. These results are used to investigate the scattering from isolated silicon and silver nanospheres.
\end{abstract}

\maketitle

The past two decades have witnessed a rapid rise of interest in the resonant electromagnetic scattering from metal and dielectric nanoparticles (NPs). Metal NPs can support coherent oscillations of their free electron plasma, also known as localized surface plasmons \cite{kreibig2013}. This phenomenon enables the fine control of electromagnetic fields at the nanoscale and the boosting of linear and nonlinear optical processes \cite{Schuller2010}, stimulating a plethora of potential applications, including ultrasensitive  biosensors \cite{anker2008biosensing}, nonlinear optics \cite{kauranen2012nonlinear}, and solar energy harvesting \cite{atwater2010plasmonics}. Recently, a new branch of nanophotonics has emerged, advocating the use of high index dielectrics and semiconductors, instead of metals \cite{Kuznetsovaag2472}, which are plagued by high losses \cite{khurgin2015deal}. 

However, despite many advancement in the understanding of NP electromagnetic scattering, several fundamental questions remain unanswered. First, the resonances in metal or dielectric NP of size comparable to the wavelength are currently found either experimentally or numerically by exciting the NP with radiation of various frequencies and locating the peaks of the corresponding scattered power spectra. However, this procedure is flawed and can be often misleading. 
Generally speaking, a peak of the scattered power spectrum cannot be univocally associated to the action of a unique {\it resonant mode} but may arise from the interplay  of several modes. The correct identification of the resonant modes is essential in the description of anomalous scattering behaviours such as the appearance of asymmetric resonant line-shapes (often referred to as Fano-like resonances \cite{luk2010fano}) where interference phenomena play a key role.
 Second, why silicon and metal NPs of comparable size exhibit deeply different resonant behaviours in the visible spectral range? For instance, magnetic modes in isolated metal nanospheres at optical frequencies have never been seen, but it is very well established that they can be  excited in their silicon counterpart \cite{kuznetsov2012magnetic}. Third, why asymmetric lineshapes in the total scattering spectrum have been observed for Si spheres, see for instance Fig. 1 (a) of Ref. \cite{Butakov16}, but not for metal spheres? 

The shortcomings of the current understanding of resonances in the electromagnetic scattering from particles can be attributed to the lack of a full electrodynamic spectral theory. In fact, the Mie theory and its extensions are not spectral theories.  In the electrostatic limit, spectral theories that allow the identification of plasmon resonances do exist. For instance, for a spherical NP of permittivity $\varepsilon_r$ and size much smaller than the incident wavelength, the Fr\"ohlich condition \cite{kreibig2013}, i.e.
\begin{equation}
     \left| \varepsilon_r \left( \omega \right) + 2 \right| = \underset{\omega}{\mbox{Minimum}}
     \label{eq:Frohlich}
\end{equation}
 enables the calculation of the resonant value of permittivity, and hence NP resonance frequency. Later, Mayergoyz et al. \cite{Fredkin2003} introduced a more general quasi-electrostatic technique applicable to arbitrarily shaped particles. However, although the applicability domain of these approaches can be reasonably extended using perturbation techniques \cite{Bohren1998,Fredkin2003}, they fail when the particle size is comparable with the incident wavelength.

\ifpaper
\begin{figure}[!t]
\centering
\includegraphics[width=\linewidth]{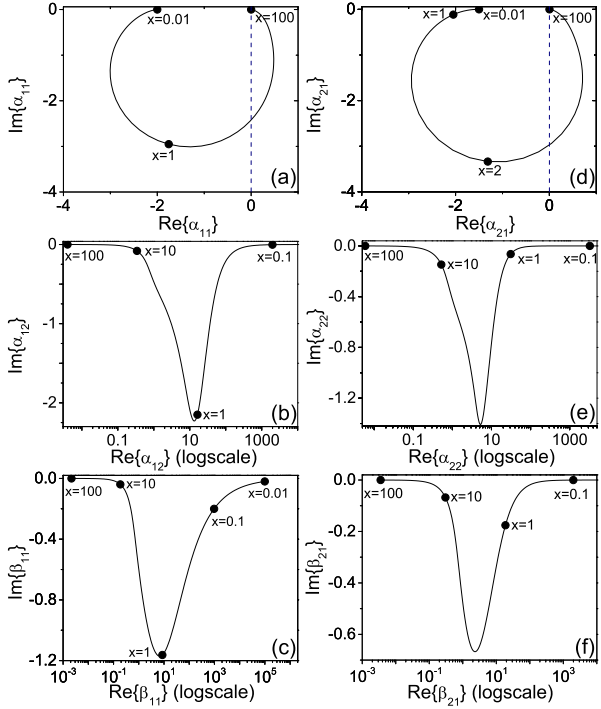}
\caption{Loci spanned in the complex plane by the poles of a homogeneous sphere by varying $x \in \left[0.01,100 \right]$. We show the loci associated to the electric dipole ($n=1$) with $l=1$ (a) and $l=2$ (b), of the electric quadrupole ($n=2$)  with $l=1$ (d) and $l=2$ (e), and of the fundamental magnetic dipole (c) and quadrupole (f).
}
  \label{fig:Loci}
\end{figure}
\fi

Here, we apply a spectral method to the calculation of the resonance frequencies of the electromagnetic scattering from spherical particles in the full-retarded
regime. It is based on a series expansion of the scattered electric field in terms of the form $\left(\gamma_h - \varepsilon_r \right)^{-1} {\bf C}_h \left( {\bf r} \right)$, where $\gamma_h$ and ${\bf C}_h$ are respectively the eigenvalues and eigenfunctions of the auxiliary eigenvalue problem introduced in Ref. \cite{Forestiere16}, which do not depend on the permittivity $\varepsilon_r$. The study of these eigenvalues and eigenfunctions unveils fundamental properties of the scattering processes, including the  difference between dielectric and metal nanoparticles due to the disjoint subset of narrow modes excited in these two cases. Similar approaches have been  introduced in the past \cite{Fuchs75,Bergman80} and applied to the quasi-static electromagnetic scattering \cite{Fuchs75,Bergman78,Bergman80,Rojas86,Fredkin2003,Mayergoyz05}, to the retarded single dipole approximation \cite{Markel:95}, to the scalar Mie scattering \cite{Markel10}, and to the full-wave electromagnetic scattering from a flat-slabs \cite{Bergman16}. 

Let us consider the electromagnetic scattering by a sphere of radius $R$ occupying a regular region $\Omega$. The sphere is excited by a time harmonic electromagnetic field incoming from infinity $\re{\Ei \left({\bf r}\right) e^{- i \omega t}}$ at wavelength $\lambda = 2\pi c / \omega$, where $c$ is the speed of light in vacuum. The material is a non-magnetic isotropic homogeneous dielectric with relative permittivity $\varepsilon_r \left( \omega \right)$, surrounded by vacuum. 

The solution of the electromagnetic scattering problem is represented as 
 \cite{Forestiere16}:
\begin{equation}
 {\bf E}_S \left( {\bf r} \right) = \left( \varepsilon_r - 1 \right)
\displaystyle\sum_{pmnl} \left(  \frac{A_{pmnl}}{ \alpha_{nl} - \varepsilon_r } \Caf{p}{m}{n}{l} \left( {\bf r} \right) + \frac{B_{pmnl}}{ \beta_{nl} - \varepsilon_r } \Cbf{p}{m}{n}{l} \left( {\bf r} \right) \right),
\label{eq:ExpansionEi}
\end{equation}
where the eigenvalues $\alpha_{nl}$, $\beta_{nl}$ and the corresponding eigenfunctions $\Caf{p}{m}{n}{l}$,  $\Cbf{p}{m}{n}{l}$ are provided in Ref. \cite{Forestiere16}, and
\begin{equation}
A_{ p m n l} = \frac{\langle 
\Caf{p}{m}{n}{l}, {\bf E}_i \rangle_\Omega}{\langle 
\Caf{p}{m}{n}{l}, \Caf{p}{m}{n}{l} \rangle_\Omega}, \qquad
B_{ p m n l} = \frac{\langle 
\Cbf{p}{m}{n}{l}, {\bf E}_i \rangle_\Omega}{\langle 
\Cbf{p}{m}{n}{l}, \Cbf{p}{m}{n}{l} \rangle_\Omega},
\label{eq:EiCoeff}
\end{equation}
\begin{equation}
\displaystyle\sum_{pnml} =
 \displaystyle\sum_{p\in \left\{e,o \right\}} \displaystyle\sum_{n=1}^\infty \displaystyle\sum_{m=0}^n  \displaystyle\sum_{l=1}^\infty, \qquad \langle \mathbf{A},\mathbf{B} \rangle_V = \iiint_V \mathbf{A} \cdot \mathbf{B} \, \mbox{dV}.
\end{equation}
The set of poles $\left\{ \alpha_{nl},  \beta_{nl} \right\}$ and the corresponding mode functions $\left\{ \Caf{p}{m}{n}{l}, \Cbf{p}{m}{n}{l} \right\}$
in Eq. \ref{eq:ExpansionEi} are complex and do not depend on the sphere permittivity $\varepsilon_r$, but only on the dimensionless parameter $x = 2 \pi R/\lambda = \omega R / c$.
Due to the Silver-M\"uller condition at infinity, the imaginary part of $\left\{ \alpha_{nl},  \beta_{nl} \right\}$ is always negative.

The modes $\Caf{p}{m}{n}{l}$ are of electric type  and the modes $\Cbf{p}{m}{n}{l}$ are of magnetic type. The modal indices $p,m,n,l$ have the following meaning.
The subscript $p \in \left\{e,o\right\}$ distinguishes between even and odd modes with respect to the azimuthal dependence. The numbers $n\in\mathbb{N}$ and $ 0 \le m \le n$ define the angular dependence of the mode: $n$ is associated to the number of lobes of the mode amplitude at any given radial distance. In particular, the modes with $n=1$ are the dipolar modes, those with $n=2$ are the quadrupolar modes, and so on.
The mode number $l\in\mathbb{N}$ gives the number of maxima of the mode amplitude along the radial direction inside the sphere. We denote the electric and magnetic modes as {\it fundamental} when $l=1$, and as {\it higher order} modes when $l\ge 2$.  In general, we have $\langle {\Cdf{p}{m}{n}{l}} \; \Cdpf{p'}{m'}{n'}{l'} \rangle_\Omega = 0$, $\forall \left(\delta,p,m,n,l\right) \ne \left(\delta',p',m',n',l'\right)$, where $\delta,\delta' \in \left\{ \alpha ,  \beta \right\}$. Furthermore, due to the spherical symmetry, we also have $\langle \left({\Cdf{p}{m}{n}{l}}\right)^* \,,\, \Cdpf{p'}{m'}{n'}{l'} \rangle_\Omega = 0$, $\forall \left(\delta,p,m,n\right) \ne \left(\delta',p',m',n'\right)$. 

The expansion \ref{eq:ExpansionEi} allow us to address the first question formulated in the introduction. 
For passive materials with $\im{\varepsilon_r} \ge 0$, the quantities $\left| \alpha_{nl} - \varepsilon_r \right|$ and $\left| \beta_{nl} - \varepsilon_r \right|$ in Eq. \ref{eq:ExpansionEi} do not vanish as $\omega$ varies for a given $R$, because $\im{\alpha_{nl}}<0$ and $\im{\beta_{nl}}<0$. Nevertheless, the mode amplitudes $ {A_{ p m n l}}/{ \left( \alpha_	{nl} - \varepsilon_r  \right) }$ and $ {B_{ p m n l}}/{ \left( \beta_{nl} - \varepsilon_r \right)}$ reach their maximum whenever for a given $R$ and $\varepsilon_r \left(x\right)$
\begin{equation}
   \begin{aligned}
   r_{nl}^{\left( \alpha \right)} = \left| \varepsilon_r \left( x \right)  - \alpha_{nl}\left( x \right) \right| &= \underset{x}{\mbox{Minimum}}; \\
   r_{nl}^{\left( \beta \right)} =  \left| \varepsilon_r \left( x \right) - \beta_{nl} \left( x \right) \right| &= \underset{x}{\mbox{Minimum}},
   \label{eq:ResonantCondition}
   \end{aligned}
\end{equation}
respectively.  These are the conditions that maximize the contribution of the modes $\Caf{p}{m}{n}{l}$ and $\Cbf{p}{m}{n}{l}$, respectively, as $x$ varies and define their resonance frequencies. The resonant width of the mode depends on the minimum values of the residua. The
mode is {\it narrow} if its minimum residuum is less than a given quantity $\rho$, {\it broad} otherwise. In the following, we assume $\rho=5$.


Since $\alpha_{nl}$ and $\beta_{nl}$ do not depend on the sphere's permittivity, their behavior can be completely described by the loci they span in the  complex plane as a function of $x$. The resulting diagrams are universal, being applicable to every possible homogeneous sphere, and constitute an invaluable tool to investigate NPs resonances.
As already noted, $\alpha_{nl}$ and $\beta_{nl}$ have negative imaginary part. Instead, their real part can be either positive or negative. When it is negative, the condition \ref{eq:ResonantCondition} is verified by metals at optical frequencies ($ \mbox{Re} \left\{ \varepsilon_r \right\} <0$), giving rise to {\it plasmon resonances}. When it is positive, the condition $\ref{eq:ResonantCondition}$ is verified by dielectrics ($ \mbox{Re} \left\{ \varepsilon_r \right\} \ge 0$), giving rise to {\it photonic resonances}. 

The analysis of the loci of $\alpha_{nl}$ and $\beta_{nl}$ is propaedeutic to the understanding of the scattering from an isolated homogeneous sphere, regardless of its size and composition. Therefore, in Fig. \ref{fig:Loci} we plot the loci spanned by $\alpha_{1l}$ for the fundamental ($l=1$) (a) and higher order ($l=2$) (b) electric dipole. In Fig. \ref{fig:Loci} (c) we plot the loci of $\beta_{11}$ for the fundamental magnetic dipole. In Fig. \ref{fig:Loci} we also plot the loci spanned by $\alpha_{2l}$ for the fundamental ($l=1$) (d) and higher order ($l=2$) (e) electric quadrupole.
In Fig. \ref{fig:Loci} (f) we plot the loci of $\beta_{21}$ of the fundamental magnetic quadrupole.

For $x \ll 1$ (Rayleigh limit), $\alpha_{11}$  approaches the value $-2$ (see Fig. \ref{fig:Loci} (a)), and the condition \ref{eq:ResonantCondition} reduces to the Fr\"{o}hlich condition \ref{eq:Frohlich}. Therefore, for finite values of $x$, the condition \ref{eq:ResonantCondition} applied to the fundamental dipole represents the natural extension of the Fr\"{o}hlich condition to the full retarded case. Both the fundamental dipole $\Caf{p}{m}{1}{1}$ and quadrupole  $\Caf{p}{m}{2}{1}$ have their corresponding pole's locus confined within a limited region of the complex plane, being $ -3 \le \re{\alpha_{11}} \le 0.48$ and $ -2.94 \le \re{\alpha_{21}} \le 0.71$. Therefore, in light of Eq. \ref{eq:ResonantCondition} they cannot be narrowly excited in particles with moderately positive permittivity such as Si because they are broad. Instead, they are narrow for metal particles in the visible spectral range, whose permittivity belongs to the third quadrant of the complex plane.

The loci spanned by $\alpha_{1l}$, $\alpha_{2l}$ with $l\ge 2$, associated to higher order electric dipolar  and quadrupolar modes, and by $\beta_{nl}$ associated to all the magnetic-type modes are profoundly different, as shown in Fig. \ref{fig:Loci} (b-c),(e-f). They always lie in the fourth quadrant  irrespectively of $x$. Moreover, they asymptotically approach the positive real axis for $x \rightarrow 0$. Thus, they cannot be narrowly excited in metal NPs
 with $\re{\varepsilon_r} \le 0$ and play no role in the scattering of any deep subwavelength particles. 

Thus, the only modes that can be narrowly excited in a metal sphere are the fundamental electric ones. These modes constitute an orthogonal set, this fact prevents interference phenomena in the total scattered power. Vice versa, the modes that can be narrowly excited in a dielectric sphere with positive permittivity are the magnetic modes and the higher order electric modes ($l\ge 2$). Some of these  modes are non-orthogonal, enabling interference phenomena. These general properties allow us to answer also to the second and third questions posed in the introduction, as we will explicitly see in the following.

We now use this approach to investigate the resonances of silicon (Si) and silver (Ag) spheres with $R=100$nm and $60$nm, respectively. We model the Ag permittivity $\varepsilon_{r,Ag}$  with experimental data \cite{jiang2016realistic}, instead for Si we use a constant permittivity, i.e. $\varepsilon_{r,Si}=16$. In Tabs. \ref{tab:ResonancesSi} and \ref{tab:ResonancesAg} we list, for  few representative modes of the considered Si and Ag NPs, the values of $x$ that minimize the residua introduced in Eq. \ref{eq:ResonantCondition}, denoted as $x_{nl}^{\left(\alpha\right)}$, $x_{nl}^{\left(\beta\right)}$, the corresponding values of the poles $\alphahat = \alphares$ and $\betahat = \betares$, and the minimum residua 
$\reshata=\resresa$, $\reshatb=\resresb$. In bold we  highlight the narrow modes. We searched for the minima in the range $x \in \left[0.01, 100 \right]$ and $x \in \left[0.19, 1.08 \right]$ for Si and Ag, respectively.

\begin{table}[htbp]
\centering
\caption{\bf Values of $x$ minimizing the residua, corresponding poles and residua for a $100$nm Si sphere.}
\begin{tabular}{c|ccc|ccc}
\hline
 $ n,l $ & ${x}_{nl}^{\left( \alpha \right)}$ & $\alphahat$ & $ \reshata $ & ${x}_{nl}^{\left( \beta \right)}$ & $\betahat$ & $ \reshatb $ \\
\hline
$1,1$ &  2.20 & 0.48 - 0.96 i & 15.6 & \bf 0.75 & \bf 16.2 - 1.07 i  & \bf 1.08 \\
$1,2$ & \bf 1.06 & \bf 16.3 - 2.15 i & \bf 2.17 &  \bf 1.55 &  \bf 15.9  - 0.96 i & \bf 0.97 \\
$2,1$ & 2.99 & 0.70 - 1.33 i & 15.3 & \bf 1.09 & \bf 16.2 - 0.21 i & \bf 0.27 \\
$2,2$ & \bf 1.38 & \bf 16.0 - 0.31 i & \bf 0.31 &  \bf 1.89 & \bf 16.0 - 0.45 i & \bf 0.45   \\
$3,1$ & 3.86
 &  1.00 - 1.67 i & 
   15.1 & \bf 1.42 & \bf  15.9 - 0.04 i & \bf 0.07
 \\
\hline
\end{tabular}
  \label{tab:ResonancesSi}
\end{table}
\begin{figure}[!t]
\centering
\includegraphics[width=\linewidth]{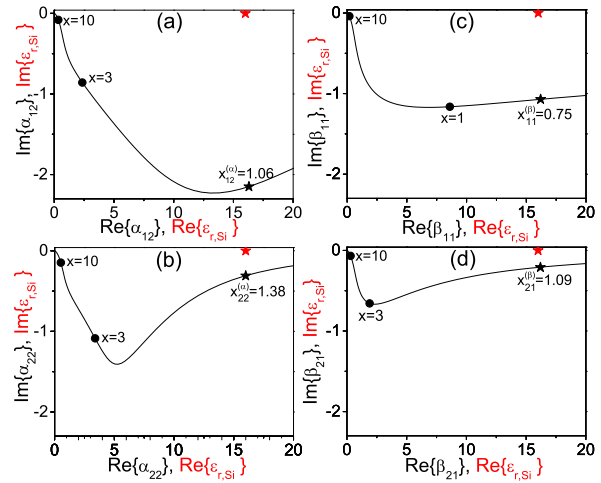}
\caption{Loci spanned in the complex plane by the poles associated to the electric dipole with $l=2$ (a) and quadrupole with $l=2$ (b) and to the fundamental magnetic dipole (c) and quadrupole (d) (black lines). Positions of the poles (black stars) $\alphahat$ with $(n,l)=$ $\left(1,2\right)$ (a), $\left(2,2\right)$ (b) and of $\betahat$  with $(n,l)=$ $\left(1,1\right)$ (a), $\left(2,1\right)$ (b). Position of $\varepsilon_{r,Si}	$ (red star).}
  \label{fig:LociSi}
\end{figure}
For the Si sphere, the residua associated to the fundamental electric modes (dipole, quadrupole, and octupole) are roughly one order of magnitude larger than the ones associated to higher order electric modes and the magnetic modes. This is consistent with the fact that the former modes are broad in a sphere with moderately positive permittivity. Furthermore, Tab. \ref{tab:ResonancesSi} shows that the real parts of the poles associated to the higher order electric modes and to the magnetic modes all approach the  value of $\varepsilon_{r,Si}=16$. This is also clear in Fig. \ref{fig:LociSi}, where we show with a red star the position of $\varepsilon_{r,Si}$, and with black stars the position of $\alphahat$  with $(n,l)=$ $\left(1,2\right)$ (a), $\left(2,2\right)$ (b) and of $\betahat$  with $(n,l)=$ $\left(1,1\right)$ (c), $\left(2,1\right)$ (d). We also plot with a black line the loci of (a) $\alpha_{12}$, (b) $\alpha_{22}$, (c) $\beta_{11}$, (d) $\beta_{21}$,  which were already examined in Fig. \ref{fig:Loci}. Fig. \ref{fig:LociSi} provides the geometrical interpretation of Eq. \ref{eq:ResonantCondition}: the black stars minimize the distance between the black curve and red star in each panel, while the distance itself represents the residuum.

\begin{table}[htbp]
\centering
\caption{\bf Values of $x$ minimizing the residua, corresponding poles and residua for a $60$nm Ag sphere.}
\begin{tabular}{c|ccc|ccc}
\hline
 $n,l$ & $ x_{nl}^{\left(\alpha \right)} $ & $\alphahat$  & $\reshata$ & $ x_{nl}^{\left(\beta \right)} $ & $\betahat$  & $\reshatb$ \\
\hline
$1,1$ & \bf 0.92 & \bf -2.42 - 2.61 i & \bf 3.0 & 1.08 & 7.6 - 1.17 i &  8.55 \\
$1,2$ & 1.25 & 11.8 - 2.2 i &  10.3 & 1.08 & 33.0 - 1.04 i &  33.9 \\
$2,1$ & \bf 1.0 & \bf -2.03 - 0.11 i & \bf 0.35 & 1.08 &  16.5 - 0.20 i & 17.4 \\
$2,2$ & 1.25 & 19.8 - 0.19 i & 17.6 & 1.08 &  50.5 - 0.19 i & 51.3 \\
\hline
\end{tabular}
  \label{tab:ResonancesAg}
\end{table}

\begin{figure}[!t]
\centering
\includegraphics[width=\linewidth]{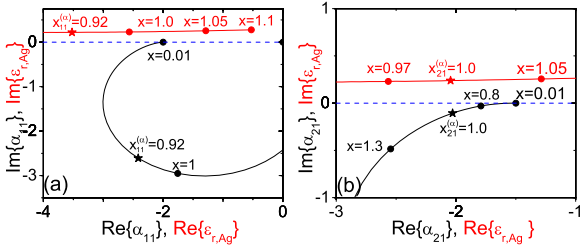}
\caption{Loci spanned in the complex plane by the poles associated to the fundamental electric dipole (a) and quadrupole (b) (black lines). Locus spanned by the permittivity $\varepsilon_{r,Ag} \left( x \right)$ of  a $60$nm Ag sphere by varying $x$ (red line). Positions of $\alphahat$ (black star) and $\varepsilon_{r,Ag} \left( x_{nl}^{\left( \alpha \right)} \right)$ (red star) for $(n,l)=\left(1,1\right)$ (a) and $(n,l)=\left(2,1\right)$ (b).}  
\label{fig:LociAg}
\end{figure}

For Ag particles exactly the opposite happens. Tab. \ref{tab:ResonancesAg} shows that the residua exhibited by all the magnetic modes and by higher order electric modes are much larger than the ones of the fundamental electric modes. Therefore, these modes are broad and play a minor role. In Fig. \ref{fig:LociAg} we  show the positions on the complex plane of $\varepsilon_{r,Si} \left( x_{nl}^{\left( \alpha \right)} \right)$ and of the pole $\alphahat$ with red and black stars, respectively, for $(n,l)=$ $\left(1,1\right)$ (a), $\left(2,1\right)$ (b).  We also plot with a black line the locus spanned as $x$ varies by $\alpha_{11}$ (a), $\alpha_{21}$ (b), and with a red line the locus of the permittivity $\varepsilon_{r,Ag}$. In both panels, the positions of the black and red stars minimize the distance between the black and red curves as $x$ varies. It is apparent from Tab. \ref{tab:ResonancesAg} and Fig. \ref{fig:LociAg} that the residuum of the fundamental dipole mode is larger than the corresponding residuum of the quadrupole. Therefore, we expect the former resonance to be broader than the latter. 
In conclusion, the modes that can be {\it narrowly} excited in metal and dielectric spheres are two disjoint sets. This fact answer the second question we posed in the introduction, justifying why silicon and metal NPs of comparable size exhibit remarkably different resonant behavior. In particular, magnetic-type modes cannot be narrowly excited in metal spheres, regardless of $x$.


\ifpaper
\begin{figure}[!t]
\centering
\includegraphics[width=\linewidth]{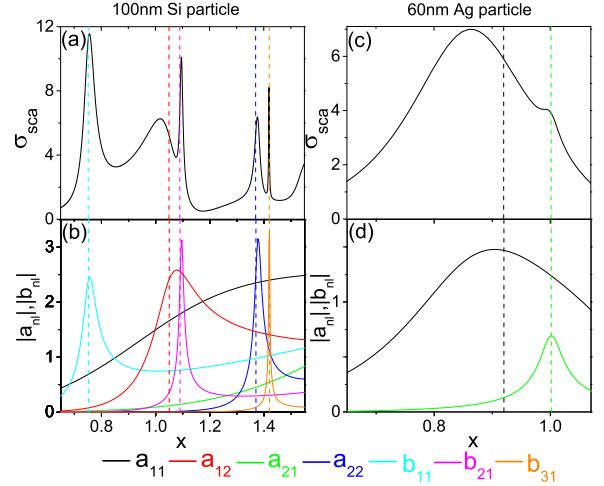}
\caption{Scattering efficiency $\sigma_{sca}$ of the investigated Si (a) and Ag (c) spheres as function of  $x$. The vertical dashed lines represent the resonance positions calculated in Tabs. \ref{tab:ResonancesSi} and \ref{tab:ResonancesAg}. Absolute value of the coefficients $a_{nl}$ and $b_{nl}$ as a function of  $x$ for Si (b) and Ag (d) spheres. }
  \label{fig:Csca}
\end{figure}
\fi

The scattering efficiency $\sigma_{sca}$ has the following expression
\begin{equation}
   \sigma_{sca} = {x^{-2}}
\sum_n  \big( \big| \sum_l a_{nl} \big|^2 + \big| \sum_l b_{nl} \big|^2 \big),
\end{equation}
where $a_{nl}= \left( { \alpha_{nl} - \varepsilon_r } \right)^{-1} \tilde{A}_{nl}$ and $b_{nl}= \left( \beta_{nl} - \varepsilon_r \right)^{-1}\tilde{B}_{nl}$. The coefficients $\tilde{A}_{nl}$, $\tilde{B}_{nl}$ do not depend on $\varepsilon_r$; they vary linearly on ${\bf E}_i$, and very slowly on  $x$. Their expression can be found in Ref. \cite{Forestiere16}. In Fig. \ref{fig:Csca} we plot $\sigma_{sca}$ of the $100$nm Si (a) and $60$nm Ag (c) sphere as a function of $x$, when excited by a linearly polarized plane wave. We also show with vertical dashed lines the resonant values of $x$ as listed in Tabs \ref{tab:ResonancesSi}, \ref{tab:ResonancesAg}. In Fig. \ref{fig:Csca} (b) and (d) we show the magnitude of the coefficients $a_{nl}$ and $b_{nl}$ for the Si and Ag sphere, respectively, as $x$ varies. 

Due to the symmetry of ${\bf E}_i$ only the modes with $m=1$ are excited. For the Si NP, all peaks but one can be attributed to the dominant contribution of a single mode. In particular, the first peak from the left is due to $\Cbf{o}{1}{1}{1}$, the third one to $\Cbf{o}{1}{2}{1}$, the fourth one to  $\Caf{e}{1}{2}{2}$, and the fifth one to $\Cbf{o}{1}{3}{1}$. Instead, the second peak  arises at $x=1.02$ from the interplay between  $\Caf{e}{1}{1}{1}$ and $\Caf{e}{1}{1}{2}$, as suggested by Fig. \ref{fig:Csca} (b).
In addition, the destructive interference between these two modes, which are  not orthogonal, is responsible for the scattering dip enclosed by peaks 2 and 3 at $x=1.076$, and of the corresponding asymmetry of $\sigma_{sca}$. In correspondence of this dip the values of the coefficients are $a_{11} = 0.78 - 0.95 i$ and $a_{12} = 0.063 + 1.83 i$. It is worth to note that, although the scattering dip is in the close proximity of the third peak, the magnetic dipole $\Cbf{o}{1}{2}{1}$  cannot be  held responsible  for it, because it is orthogonal to both $\Caf{e}{1}{1}{1}$ and $\Caf{e}{1}{1}{2}$, and interference cannot take place. We now explore the role played by the fundamental electric dipole. Although $\Caf{e}{1}{1}{1}$ is always broad in a Si sphere, it is still able to significantly contribute to $\sigma_{sca}$, because the coefficient $A_{11}$ is very large compared to $A_{nl}$ and $B_{nl}$ of the remaining modes. This is due to the fact that the fundamental dipole $\Caf{e}{1}{1}{1}$  more easily couples with the exciting plane wave and more strongly radiates into the far field.


Next, we analyse the $\sigma_{sca}$ spectrum of the Ag NP. It can be exhaustively described by considering only the fundamental electric dipole and quadrupole modes, namely $\Caf{e}{1}{1}{1}$ and $\Caf{e}{1}{2}{1}$, because the higher order electric modes and the magnetic modes cannot be narrowly excited in metal spheres. The peak associated to the mode $\Caf{e}{1}{1}{1}$ is very broad due to large imaginary part of the pole $\alphahat$, as shown in Tab. \ref{tab:ResonancesAg} and in Fig. \ref{fig:LociAg} (a). Moreover, there are no asymmetries in the resonances, due to the orthogonality of the fundamental electric modes.



\begin{thebibliography}{10}
\newcommand{\enquote}[1]{``#1''}

\bibitem{kreibig2013}
U.~Kreibig and M.~Vollmer, \emph{Optical properties of metal clusters}, vol.~25
  (Springer Science \& Business Media, 2013).

\bibitem{Schuller2010}
J.~A. Schuller, E.~S. Barnard, W.~Cai, Y.~C. Jun, J.~S. White, and M.~L.
  Brongersma, Nat. Mater. \textbf{9}, 193 (2010).

\bibitem{anker2008biosensing}
J.~N. Anker, W.~P. Hall, O.~Lyandres, N.~C. Shah, J.~Zhao, and R.~P. Van~Duyne,
  Nature materials \textbf{7}, 442 (2008).

\bibitem{kauranen2012nonlinear}
M.~Kauranen and A.~V. Zayats, Nature Photonics \textbf{6}, 737 (2012).

\bibitem{atwater2010plasmonics}
H.~A. Atwater and A.~Polman, Nature materials \textbf{9}, 205 (2010).

\bibitem{Kuznetsovaag2472}
A.~I. Kuznetsov, A.~E. Miroshnichenko, M.~L. Brongersma, Y.~S. Kivshar, and
  B.~Luk{\textquoteright}yanchuk, Science \textbf{354} (2016).

\bibitem{khurgin2015deal}
J.~B. Khurgin, Nature nanotechnology \textbf{10}, 2 (2015).

\bibitem{luk2010fano}
B.~Luk'yanchuk, N.~I. Zheludev, S.~A. Maier, N.~J. Halas, P.~Nordlander,
  H.~Giessen, and C.~T. Chong, Nature materials \textbf{9}, 707 (2010).

\bibitem{kuznetsov2012magnetic}
A.~I. Kuznetsov, A.~E. Miroshnichenko, Y.~H. Fu, J.~Zhang, and B.~Luk'Yanchuk,
  Scientific Reports \textbf{2}, 492 (2012).

\bibitem{Butakov16}
N.~A. Butakov and J.~Schuller, Scientific Reports \textbf{6}, 38487 (2016).

\bibitem{Fredkin2003}
D.~R. Fredkin and I.~D. Mayergoyz, Phys. Rew. Letters \textbf{91} (2003).

\bibitem{Bohren1998}
C.~F. Bohren and D.~R. Huffman, \emph{Absorption and Scattering of Light by
  Small Particles} (Wiley, 1998).

\bibitem{Forestiere16}
C.~Forestiere and G.~Miano, Phys. Rev. B \textbf{94}, 201406 (2016).

\bibitem{Fuchs75}
R.~Fuchs, Phys. Rev. B \textbf{11}, 1732 (1975).

\bibitem{Bergman80}
D.~J. Bergman and D.~Stroud, Physical Review B \textbf{22}, 3527 (1980).

\bibitem{Bergman78}
D.~J. Bergman, Physics Reports \textbf{43}, 377 (1978).

\bibitem{Rojas86}
R.~Rojas and F.~Claro, Phys. Rev. B \textbf{34}, 3730 (1986).

\bibitem{Mayergoyz05}
I.~Mayergoyz, D.~Fredkin, and Z.~Zhang, Phys. Rev. B \textbf{72}, 155412
  (2005).

\bibitem{Markel:95}
V.~A. Markel, J. Opt. Soc. Am. B \textbf{12}, 1783 (1995).

\bibitem{Markel10}
V.~A. Markel, Journal of Nanophotonics \textbf{4}, 041555 (2010).

\bibitem{Bergman16}
A.~Farhi and D.~J. Bergman, Physical Review A \textbf{93}, 063844 (2016).

\bibitem{jiang2016realistic}
Y.~Jiang, S.~Pillai, and M.~A. Green, Scientific Reports \textbf{6} (2016).

\end{thebibliography}
\end{document}